\documentclass[10pt,twocolumn,twoside]{IEEEtran}
\IEEEoverridecommandlockouts                            

\usepackage{cite}
\usepackage{amsmath,amssymb,amsfonts}
\usepackage{mathtools}
\usepackage{algorithmic}
\usepackage{graphicx}
\usepackage{textcomp}
\usepackage{balance}
\usepackage{xcolor}
\usepackage[utf8]{inputenc}
\usepackage{xurl}
\usepackage{booktabs} 

\usepackage{subfigure}
\usepackage[hidelinks]{hyperref}
\usepackage{multirow}

\newcommand{\hquad}{\hspace{0.5em}}
\newcommand{\norm}[1]{\left\lVert#1\right\rVert}
\newcommand{\minimize}[1]{\underset{{#1}}{\text{minimize}}}

\newcommand{\st}{\text{subject to}}
\newcommand{\mb}[1]{\mathbf{#1}}
\newcommand{\mbg}[1]{\boldsymbol{#1}}
\newcommand{\dif}{\mathrm{d}}

\newcommand{\rms}{\text{\tiny RMS}}

\newcommand{\nom}{\text{n}}
\newcommand{\tmax}{\text{max}}

\def\BibTeX{{\rm B\kern-.05em{\sc i\kern-.025em b}\kern-.08em
    T\kern-.1667em\lower.7ex\hbox{E}\kern-.125emX}}
\usepackage{lipsum}

\title{\LARGE \bf 
Online Feedback Optimization of Energy Storage \\ to Smooth Data Center Grid Impacts}

\author{Yanyong Mao, Johanna L. Mathieu, Vladimir Dvorkin%
    \vspace{-0.5cm}%
    \thanks{ 
        Yanyong Mao, Johanna L. Mathieu, and Vladimir Dvorkin are with the Department of Electrical Engineering and Computer Science, University of Michigan, Ann Arbor, MI 48109 USA. Email: {\tt kruxious@umich.edu; jlmath@umich.edu; dvorkin@umich.edu}}
}

\begin{document}
\begingroup
\allowdisplaybreaks

\maketitle

\begin{abstract}

The growing electricity demand of AI data centers introduces significant voltage variability in power networks, affecting not only their own operation but also the experience of all users sharing the network. To smooth data center impacts on power networks, we develop an online feedback optimization approach that controls distributed battery energy storage systems to mitigate voltage issues induced by data center operations. The controller adjusts the active and reactive power setpoints of distributed battery systems in response to voltage measurements, with a two-fold objective: managing voltage to minimize the magnitude of constraint violations and smoothing voltage profiles. Control performance is evaluated in a high-fidelity simulation environment that integrates a three-phase distribution feeder and a detailed battery system model, and benchmarked against a local control approach with similar objectives but without optimality guarantees and constraint enforcement. We show that the proposed controller delivers consistent voltage regulation in the long term, while the local control approach pursues the objectives more aggressively but quickly hits the storage limits.
\end{abstract}

\begin{IEEEkeywords}
Data centers, energy storage, feedback control, online optimization, voltage smoothing.
\end{IEEEkeywords}

\section{Introduction}
The popularization of artificial intelligence (AI) and the growing use of large language models (LLMs) has led to the rapid expansion of data centers, resulting in an increase in the interconnection of (very) large loads~\cite{IEA_energy_2025,li_unseen_2024,choukse_power_2025,ai-grid-impact-arxiv25}. Depending on the scale, data centers can be connected to subtransmission or distribution networks~\cite{garg2025considerations}. Unfortunately, data center operations can lead to voltage issues, affecting the users sharing the network. They also affect grid operations as voltage-sensitive loads: in July 2024, a minor fault on a 230 kV line in Virginia caused a 3-second voltage fluctuation, which triggered an automatic disconnection of 60 data centers simultaneously. The sudden loss of 1.5 GW of load caused the grid frequency to spike from 60 Hz to 61.2 Hz, nearly causing a cascading blackout across the Mid-Atlantic region~\cite{nerc_incident_2025}.

Data centers in distribution feeders present a distinct challenge for utilities. Although data center loads exhibit power factors of 95\% and above~\cite{mitchell-jackson_data_2003}, the relatively high resistance-to-reactance ratio of distribution lines~\cite{IEEE_test} means that active power variations can significantly impact distribution voltage. Traditional voltage control via substation transformer tap changing is impractical for this application: active power variability occurs on second-to-minute timescales, whereas safe tap-changer operation is limited to around 20 times per day to prevent mechanical wear and tear~\cite{msenergy_tapchanger}.

Battery energy storage systems are capable of fast response times. Certain battery system technologies are able to ramp up to their maximum output power within 50 milliseconds~\cite{feehally_battery_2016}, which aligns with the timescale of data center load fluctuations. Increasingly, battery systems are actively engaged in distribution system control, thanks to the roll out of behind-the-meter storage programs promoted by utilities, including ConnectedSolutions by National Grid~\cite{web_connectedsolutions}, Wattsmart Battery Program by Rocky Mountain Power~\cite{web_wattsmart}, and BYOD by Green Mountain Power~\cite{web_BYOD}. These programs either install batteries on the customer end, or incentivize customers to enroll their already-installed systems in programs that enable the utility to control battery aggregations as virtual power plants. This provides an opportunity to use battery systems to smooth-out grid impacts of data centers requiring the development of control algorithms that operationalize these resources.

Battery systems can help to mitigate distribution voltage problems through various control strategies, each with distinct trade-offs. Stochastic optimal power flow optimization ensures system-wide optimality but requires accurate network models and can lead to significant computational burden for large feeders~\cite{mieth2018data}. Model predictive control~\cite{8463580} incorporates constraints and real-time feedback but also relies on the accuracy of grid models and load forecasts. Traditional local control schemes, such as Volt/VAR and Volt/Watt control rules, offer computational simplicity and fast response but operate on local measurements alone, lacking coordination and optimality guarantees. Online feedback optimization (OFO) provides a middle-ground solution~\cite{bernstein2019real,colot2024optimal,ortmann_experimental_2020}: it achieves near-optimal, system-wide voltage regulation while requiring only real-time voltage magnitude measurements and approximate sensitivity information (e.g., network reactance), enabling semi-decentralized implementation without full model knowledge. We refer to~\cite{hauswirth_optimization_2024} for the comprehensive overview of online feedback optimization.

Our contribution is the development of a tractable controller for battery systems that mitigates data center voltage impacts on distribution feeders. Our technical contributions include:
\begin{enumerate}
    \item We propose a OFO-based voltage controller for power distribution grids that leverages customer-side batteries to smooth-out data center impacts. The controller exploits linearized power flow and energy storage models to enable real-time implementation while requiring only voltage and energy storage state of charge (SoC) measurements and approximate network parameters.
    It is designed to target two important objectives: maintain voltage magnitude within limits, and \textit{smooth} voltage profiles.  
    \item To assess the added value of the OFO controller for distribution grid operations, we benchmark it against a local control approach with comparable objectives. The benchmark controller fairs well as compared to OFO in the short-run, but fails to perform consistently on longer horizons due to lack of explicit constraint enforcement and optimality guarantees. 
    \item We develop a comprehensive testbed for numerical simulations, integrating an OpenDSS~\cite{dugan_opendss} simulator of a three-phase IEEE feeder, an accurate battery system model that accounts for energy losses in providing both active and reactive power capabilities, and the real traces of GPU power consumption during inference on Llama-3 model. 
\end{enumerate}

The remainder of this paper is organized as follows: Sec.~\ref{sec:problem} introduces the master problem which formulates the desired objectives and constraints for smoothing data center voltage impacts, yet remains computationally infeasible in practice. In Sec.~\ref{sec:OFO}, we propose a tractable online solution to the master problem, specifically, the OFO controller. Sec.~\ref{sec:benchmark} details a simpler benchmark PI controller with similar objectives, and Sec.~\ref{sec:case-study} introduces the testbed and reports numerical results. Finally, Sec.~\ref{sec:conclusion} concludes the paper.

\section{Problem Definition} \label{sec:problem}

Consider the problem of smoothing out the voltage impacts of AI data centers on distribution systems using several storage systems. The impacts are characterized by the voltage limit violations (e.g., voltage magnitudes outside the standard $\pm0.05$ p.u. range) and intertemporal voltage fluctuations. We aim to devise a controller that regulates active and reactive power setpoints of distributed battery systems in response to voltage and SoC measurements. The integrated storage and distribution systems can thus be regarded as the plant, and data center load as the disturbance to the plant. When implemented in real-time, the proposed controller takes measurement from the plant as feedback, makes control decisions and then sends control signals to battery systems, as illustrated in Fig.~\ref{fig:system-diagram}.
\begin{figure}[t]
    \centering
    \includegraphics[width=1.0\linewidth]{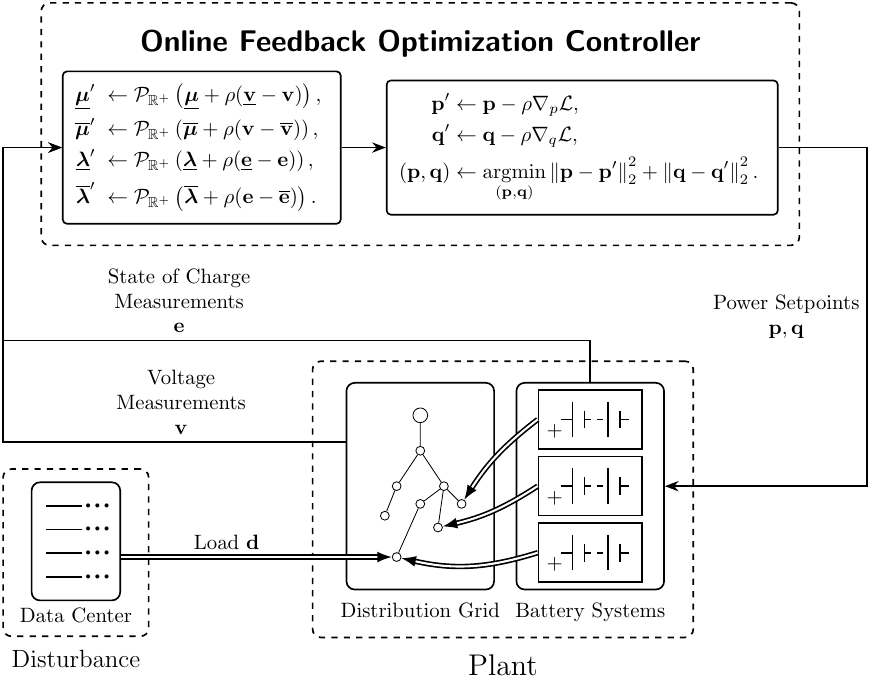}
    \caption{The scope of the system. Battery energy storage systems and the power distribution system form the plant. The OFO controller takes bus voltage and battery SoC measurements from the plant, and determines the power setpoints for battery systems. Load fluctuations of AI data centers are seen as disturbances to the plant.}
    \label{fig:system-diagram}
\end{figure}

The master optimization problem that formulates these control objectives and constraints is as follows:
\begin{subequations}\label{prob:master}
    \begin{align}
        \minimize{\substack{\mb{v}, \mb{e}, \mb{p}, \mb{q}}} \hquad & \sum_{t=1}^{T} \big( c_v(\mb{v}_t, \mb{v}_{t-1})  + c_{pq}(\mb{p}_t, \mb{q}_t) \big)\label{eq:master-objective}\\
        \st \hquad
        & \mb{v}_t = f(\mb{p}_t,\mb{q}_t, \mb{d}_t), \hquad \forall t \in \mathcal{T}, \label{eq:master-grid-model}\\
        & \underline{\mb{v}} \leqslant \mb{v}_{t} \leqslant \overline{\mb{v}}, \hquad \forall t \in \mathcal{T}, \; \label{eq:master-voltcon}\\
        & \mb{p}_{t}^2 + \mb{q}_{t}^2 \leqslant \overline{\mb{s}}^2, \quad \forall t \in \mathcal{T}, \label{eq:master-powercon}\\
        & \mb{e}_{t+1} = g(\mb{p}_t,\mb{q}_t, \mb{e}_{t}), \hquad \forall t \in \{0\} \cup \mathcal{T} \setminus \{T\}, \label{eq:master-socdynm}\\
        & \underline{\mb{e}} \leqslant \mb{e}_t \leqslant \overline{\mb{e}}, \quad \forall t \in \mathcal{T}, \label{eq:master-soccon}\\
        & \mb{e}_{T} = \mb{e}_0, \label{eq:master-socdrift}
    \end{align}
\end{subequations}
where \(\mathcal{T}=\{1, 2, \dots, T\}\) is the time horizon; subscript \(t\) indexes the time step; \(\mb{v}\) is the vector of voltage magnitudes across all buses and all phases present at each bus; \(\mb{p}\) and \(\mb{q}\) are vectors of active and reactive power injections across all phases to which each battery is connected; \(\mb{d}\) is the vector of complex power demand from both existing loads and data centers at each bus and all phases that are present at each bus; \(\overline{\mb{v}}, \underline{\mb{v}}\) are vectors of the upper and lower voltage magnitude limits; \(\overline{\mb{s}}\) is the vector of battery apparent power limits; \(\mb{e}\) is the vector of SoC levels across all batteries; and \(\overline{\mb{e}}, \underline{\mb{e}}\) are the vectors of upper and lower SoC limits. All inequalities involving vectors and square operation of vectors in this paper should be interpreted as element-wise.

The objective \eqref{eq:master-objective} consists of the sum of two cost functions: \(c_v\) penalizes voltage fluctuations, voltage limit violations, or other voltage metrics; while \(c_{pq}\) minimizes the control effort. Constraint \eqref{eq:master-grid-model} represents the model of the network, e.g., the AC power flow equations, showing that bus voltages are functions of battery power injections and (data center) demand; \eqref{eq:master-voltcon} requires bus voltages to remain within their limits; \eqref{eq:master-powercon} ensures batteries are operating within their apparent power limits; \eqref{eq:master-socdynm} models the SoC transition between two time steps; \eqref{eq:master-soccon} enforces SoC limits; and \eqref{eq:master-socdrift} restricts the SoC to take the same value at the beginning and end of each horizon. We implicitly assume the feeder is voltage-constrained and not flow-constrained, and so we do not include power flow limits.

\section{OFO Controller Design} \label{sec:OFO}
The master problem \eqref{prob:master} is computationally expensive, and requires full knowledge of the system, such as load profiles across the time horizon, accurate grid and storage (nonconvex) models, etc. Although solving it would give the optimal result, in practice any model mismatch and/or forecast error will translate into suboptimal performance compared to what could have been achieved with perfect model information. To address these challenges, we propose to solve the master problem by relying on real-time measurements of voltage and SoC as opposed to using their models. Using measurements, we can reduce the computational burden and suboptimality as compared to traditional model-based optimization techniques.

\subsection{Online Feedback Optimization}

OFO is well-suited for this scenario, because it can be readily derived from the master problem \eqref{prob:master}. Its feedback nature eliminates the need to obtain or predict the exact future load profiles, and it does not need perfect models of battery systems and grid.
Acting solely on the current plant state and model approximations, the OFO solutions can differ from the actual optimal state of \eqref{prob:master}, yet integrating its constraints and objectives it attempts to track the optimal state online. 

To formulate the tractable OFO problem, we define \(c_v\) as measure of voltage fluctuations using a weighted \(\ell^2\)-norm on the difference between the current and previous voltage measurements. For the cost of control effort, we use a convex quadratic function in active and reactive power injections. The expressions for the cost function in \eqref{eq:master-objective} take the form:
\begin{subequations}
    \begin{align}
        c_v &\coloneq \tfrac{1}{2}\norm{\mb{v}(\mb{p},\mb{q}) - \mb{v}_{\text{prev}}}_{\mb{C}_\text{vf}}^2, \\
        c_{pq} &\coloneq \tfrac{1}{2} \norm{\mb{p}}_{\mb{C}_\text{p}}^2 + \tfrac{1}{2} \norm{\mb{q}}_{\mb{C}_\text{q}}^2,
    \end{align}
\end{subequations}
where notation \(\norm{\mb{x}}_{\mb{C}}^2 = \mb{x}^\top \mb{C} \mb{x}\). Here, \(\mb{v}(\mb{p},\mb{q})\) is the vector of \textit{squared} voltage magnitude measurements (this modeling choice will be useful given the power flow linearization model used in the next subsection), and \(\mb{v}_\text{prev}\) is the vector of squared voltage magnitude measurements at the previous time step. Matrices \(\mb{C}_{\text{vf}}, \mb{C}_{\text{p}}, \mb{C}_{\text{q}} \succ 0\) collect the weights for each cost component.

Formulating OFO constraints, we discard the grid and storage models $f$ and $g$, respectively, as they are replaced by measurements. The OFO problem at each time step $t$ then takes the form: 
\begin{subequations} \label{eq:prob-ofo}
    \begin{align}
        \hspace{-0.4em}\minimize{\mb{p}, \mb{q}} \hquad
        & \tfrac{1}{2}\norm{\mb{v}(\mb{p},\mb{q}) - \mb{v}_{\text{prev}}}_{\mb{C}_\text{vf}}^2 
        + \tfrac{1}{2} \norm{\mb{p}}_{\mb{C}_\text{p}}^2 + \tfrac{1}{2} \norm{\mb{q}}_{\mb{C}_\text{q}}^2 \label{eq:ol-objective}\\[0.5em]
        \hspace{-0.4em}\st \hquad
        & \underline{\mb{v}} \leqslant \mb{v}(\mb{p},\mb{q}) \leqslant \overline{\mb{v}} \hquad : \underline{\mbg{\mu}}, \overline{\mbg{\mu}}, \label{eq:ol_voltcon}\\
        & \underline{\mb{e}} \leqslant \mb{e}(\mb{p},\mb{q}) \leqslant \overline{\mb{e}} \hquad \;: \underline{\mbg{\lambda}}, \overline{\mbg{\lambda}}, \label{eq:ol_soccon}\\
        & \mb{p}^2 + \mb{q}^2 \leqslant \overline{\mb{s}}^2, \label{eq:ol-powercon} 
    \end{align}
\end{subequations}
where we have dropped the time index $t$ for concision.

The six constraints in the master problem are reduced to three thanks to the measurements $\mb{v}$ and $\mb{e}$ replacing the models. The dual variables of the voltage and SoC constraints are stated after the colons. We intend to measure violations of these constraints and resolve. The set of decision variables is reduced to those we have direct control over. Voltage and SoC measurements are functions of controllable variables \(\mb{p}, \mb{q}\) given by plant physics, but they are no longer computed through models including auxiliary decision variables. Solving a quadratic convex optimization problem \eqref{eq:prob-ofo} at every time step allows us to (approximately) track the optimal solution of the master problem \eqref{prob:master} without dealing with its large computational overhead.

\subsection{Controller Update Policy}
To solve the OFO problem \eqref{eq:prob-ofo}, we adopt a mixed saddle-point flow dynamics approach~\cite{hauswirth_optimization_2024}. Consider the partial Lagrangian function of \eqref{eq:prob-ofo} dualizing the voltage and SoC constraints only:
\begin{align}
    \mathcal{L}(\mb{p}, \mb{q}, \underline{\mbg{\mu}}, \overline{\mbg{\mu}}, \underline{\mbg{\lambda}}, \overline{\mbg{\lambda}}) =&\; \tfrac{1}{2}\norm{\mb{v} - \mb{v}_{\text{prev}}}_{\mb{C}_\text{vf}}^2 + \tfrac{1}{2} \norm{\mb{p}}_{\mb{C}_\text{p}}^2 + \tfrac{1}{2} \norm{\mb{q}}_{\mb{C}_\text{q}}^2 \nonumber\\
    &+ \underline{\mbg{\mu}}^\top \left( \underline{\mb{v}} - \mb{v} \right) + \overline{\mbg{\mu}}^\top \left( \mb{v} - \overline{\mb{v}} \right) \nonumber\\
    &+ \underline{\mbg{\lambda}}^\top \left( \underline{\mb{e}} - \mb{e} \right) + \overline{\mbg{\lambda}}^\top \left( \mb{e} - \overline{\mb{e}} \right)
    .\label{eq:part_lag}
\end{align}
The optimization dynamics seeks the saddle point of \eqref{eq:part_lag} over discrete time steps, where the primal variables, i.e., control decisions, conduct gradient descent on \eqref{eq:part_lag}, and the dual variables are updated using projected dual ascent on \eqref{eq:part_lag}. Specifically, the primal update requires the following gradients:
\begin{subequations} \label{eq:update-incre-primal}
    \begin{align}
        \nabla_{\mb{p}} \mathcal{L} &=\; \left( \frac{\partial \mb{v}(\mb{p}, \mb{q})}{\partial\mb{p}} \right)^\top (\mb{v} - \mb{v}_{\text{prev}}) + \mb{C}_{\text{p}}^\top \mb{p} \nonumber\\
        + & \left( \frac{\partial \mb{v}(\mb{p}, \mb{q})}{\partial\mb{p}} \right)^\top\!\big( \overline{\mbg{\mu}} - \underline{\mbg{\mu}} \big)  \!+\! \left( \frac{\partial \mb{e}(\mb{p}, \mb{q})}{\partial\mb{p}} \right)^\top\! \big( \overline{\mbg{\lambda}}\!-\!\underline{\mbg{\lambda}} \big), \\
        \nabla_{\mb{q}} \mathcal{L} &=\; \left( \frac{\partial \mb{v}(\mb{p}, \mb{q})}{\partial\mb{q}} \right)^\top (\mb{v} - \mb{v}_{\text{prev}}) + \mb{C}_{\text{q}}^\top \mb{q} \nonumber\\
        + & \left( \frac{\partial \mb{v}(\mb{p}, \mb{q})}{\partial\mb{q}} \right)^\top\!\big( \overline{\mbg{\mu}} - \underline{\mbg{\mu}} \big)\!+\!\left(\frac{\partial \mb{e}(\mb{p}, \mb{q})}{\partial\mb{q}} \right)^\top\!\big( \overline{\mbg{\lambda}}\!-\!\underline{\mbg{\lambda}} \big). 
    \end{align}
\end{subequations}
The gradient direction for active and reactive power setpoints in \eqref{eq:update-incre-primal} is determined by several variables. First are the sensitivities of voltage and SoC variables to battery power injections. Second are the intertemporal voltage fluctuations. Last are the dual variables, whose values are non-zero when the associated constraints are violated. The dual variables, in turn, are updated using the following gradients:
\begin{subequations} \label{eq:update-incre-dual}
\begin{align}
\nabla_{\underline{\mbg{\mu}}} \mathcal{L} =&\;  \underline{\mb{v}} - \mb{v}, \quad \nabla_{\overline{\mbg{\mu}}} \mathcal{L} =\; \mb{v} - \overline{\mb{v}},\\
\nabla_{\underline{\mbg{\lambda}}} \mathcal{L} =&\; \underline{\mb{e}} - \mb{e}, \;\;\!\!\quad \nabla_{\overline{\mbg{\lambda}}} \mathcal{L} =\; \mb{e} - \overline{\mb{e}},
\end{align}
\end{subequations}
all reflecting the extent of the associated constraint violation.

\begin{figure*}
\centering
\begin{minipage}{\textwidth}
\normalsize
\hrule
\vspace{-0.2em}
\begin{subequations} \label{eq:primal-updates}
    \begin{align}
        \mb{p}^\prime \; \leftarrow \; & \mb{p} - \rho \Big (
          \overbrace{\vphantom{\dfrac{1}{2}} \mb{R}_{3\phi}^\top (\mb{v} - \mb{v}_{\text{prev}})}^{\text{voltage smoothing}}
        + \overbrace{\vphantom{\dfrac{1}{2}} \mb{C}_{\text{p}}^\top \mb{p}}^{\substack{\text{control} \\ \text{effort}}}
        + \overbrace{\vphantom{\dfrac{1}{2}} \mb{R}_{3\phi}^\top \big( \overline{\mbg{\mu}} - \underline{\mbg{\mu}} \big) }^{\substack{\text{voltage limit} \\ \text{ violation}}}
        + \overbrace{\vphantom{\dfrac{1}{2}} \mathrm{diag} \Big(\big( \frac{1}{\eta^-} \mathrm{sgn} (\mb{p}^-) - \eta^+ \mathrm{sgn} (\mb{p}^+) \big) \Delta t \Big)^\top \big( \overline{\mbg{\lambda}} - \underline{\mbg{\lambda}} \big)}^{\text{SoC limit violation}}
       \Big ), \\[-0.25em]
        \mb{q}^\prime \; \leftarrow \; & \mb{q} - \rho \Big(
          \mb{X}_{3\phi}^\top (\mb{v} - \mb{v}_{\text{prev}})
        + \mb{C}_{\text{q}}^\top \mb{q}
        + \mb{X}_{3\phi}^\top \big( \overline{\mbg{\mu}} - \underline{\mbg{\mu}} \big)
        + \mathrm{diag} \Big( \frac{1 - \eta^+ \eta^-}{\pi \eta^+} \mathrm{sgn} (-\mb{q}^+ - \mb{q}^-) \Delta t \Big)^\top \big( \overline{\mbg{\lambda}} - \underline{\mbg{\lambda}} \big)
        \Big), \\[0.2em]
        (\mb{p}, \mb{q}) \; \leftarrow \; & \underset{(\mb{p}, \mb{q})}{\mathrm{argmin}} \hquad \norm{\mb{p} - \mb{p}^\prime}_2^2 + \norm{\mb{q} - \mb{q}^\prime}_2^2 \label{eq:primal-projection}\\[-0.2em]
        &\hspace{-1em} \st \quad \mb{p}^2 + \mb{q}^2 \leqslant \overline{\mb{s}}^2 \nonumber.
    \end{align}
\end{subequations}
\vspace{-0.8em}
\hrule
\end{minipage}
\end{figure*}

The sensitivities of voltage and SoC variables to battery power injections can be estimated from reasonable models of these variables. For example, using LinDist3Flow~\cite{taheri_optimizing_2025} (described in Appendix A), the sensitivity of voltages with respect to active and reactive setpoints is given by matrices $\mb{R}_{3\phi}$ and $\mb{X}_{3\phi}$ incorporating line resistances, reactances, and inter-phase coupling. Model errors are likely to affect the magnitude of the gradient, but not the main directions. This property is known as the robustness of feedback optimization to model mismatch~\cite{ortmann_experimental_2020}. We derive the following sensitivities using the grid and storage models detailed in Appendix~\ref{sec:grid-model}--\ref{sec:storage-model}:%
\begin{subequations} \label{eq:sensitivities}
    \begin{align}
        \frac{\partial \mb{v}(\mb{p}, \mb{q})}{\partial\mb{p}} &= \mb{R}_{3\phi}, \quad \frac{\partial \mb{v}(\mb{p}, \mb{q})}{\partial\mb{q}} = \mb{X}_{3\phi}, \\
        \frac{\partial \mb{e}(\mb{p}, \mb{q})}{\partial\mb{p}} &= \mathrm{diag} \left(\!\left( \frac{1}{\eta^-} \mathrm{sgn} (\mb{p}^-) - \eta^+ \mathrm{sgn} (\mb{p}^+) \!\right) \!\Delta t \right), \\
        \frac{\partial \mb{e}(\mb{p}, \mb{q})}{\partial\mb{q}} &= \mathrm{diag} \bigg( \frac{1 - \eta^+ \eta^-}{\pi \eta^+} \mathrm{sgn} (-\mb{q}^+ - \mb{q}^-) \Delta t \bigg),
    \end{align}
\end{subequations}
where \(\mathrm{sgn}(\cdot)\) is the signum function outputting \(+1, 0 \text{ or } -1\) based on the sign of the input; \(\mathrm{diag}(\cdot)\) is the diagonal operator that creates a diagonal matrix based on the input vector; \(\Delta t\) is the time step length; power setpoints are split into positive parts \(\mb{p}^+, \mb{q}^+\) (injections) and negative parts \(\mb{p}^-, \mb{q}^-\) (absorptions) due to their different effects on SoC dynamics, and \(\eta^+\) and \(\eta^-\) are the discharging and charging efficiency, respectively.

With these sensitivities in hand, the OFO controller is implemented as illustrated in Fig. \ref{fig:system-diagram}. At each time step, the controller first updates the dual variables using voltage and SoC measurements:
\begin{subequations} \label{eq:dual-updates}
    \begin{align}
        \underline{\mbg{\mu}}^\prime \; \leftarrow \; & \mathcal{P}_{\mathbb{R}^+} \left( \underline{\mbg{\mu}} + \rho (\underline{\mb{v}} - \mb{v}) \right), \\
        \overline{\mbg{\mu}}^\prime \; \leftarrow \; & \mathcal{P}_{\mathbb{R}^+} \left( \overline{\mbg{\mu}} + \rho (\mb{v} - \overline{\mb{v}}) \right), \\
        \underline{\mbg{\lambda}}^\prime \; \leftarrow \; & \mathcal{P}_{\mathbb{R}^+} \left( \underline{\mbg{\lambda}} + \rho (\underline{\mb{e}} - \mb{e}) \right), \\
        \overline{\mbg{\lambda}}^\prime \; \leftarrow \; & \mathcal{P}_{\mathbb{R}^+} \left( \overline{\mbg{\lambda}} + \rho (\mb{e} - \overline{\mb{e}}) \right),
    \end{align}
\end{subequations}
where \(\rho>0\) is the user-selected gradient assent step size, and \(\mathcal{P}_\mathbb{U} (\cdot)\) is the element-wise operator that projects its argument onto a given set \(\mathbb{U}\). The dual variable update is followed by the primal update of controllable battery injections via \eqref{eq:primal-updates}. Primal variables are first updated by taking the gradient descent with the same step size \(\rho\), and then projecting the resultant setpoints onto the inverter's feasible region.

\subsection{Anti-windup Policy}
Violations of constraints \eqref{eq:ol_voltcon} and \eqref{eq:ol_soccon} accumulate on their dual variables, making them behave like integrators. When the battery system is saturated before voltage constraints are satisfied, the dual variables grow without affecting the control action, leading to sluggish or even oscillatory behavior as errors unwind \cite{astrom_windup}. To prevent this, we designed a simple anti-windup mechanism:
\begin{subequations}
\begin{align}
    \overline{\mbg{\mu}} &= 
        \overline{\mbg{\mu}}_{\text{prev}}, \hquad \text{if } \mb{p}^2 + \mb{q}^2 = \overline{\mb{s}}^2 \text{ and } \mb{p} < 0, \mb{q} < 0, \\
    \underline{\mbg{\mu}} &= 
        \underline{\mbg{\mu}}_{\text{prev}}, \hquad \text{if } \mb{p}^2 + \mb{q}^2 = \overline{\mb{s}}^2 \text{ and } \mb{p} > 0, \mb{q} > 0,
\end{align}
\end{subequations}
where \(\overline{\mbg{\mu}}_{\text{prev}}\), \(\underline{\mbg{\mu}}_{\text{prev}}\) are the dual variables before the last update.

This mechanism checks the saturation status of battery systems. If true, it discards the latest dual update. Here, we do not consider dual variables \(\overline{\mbg{\lambda}}\) and \(\underline{\mbg{\lambda}}\) of SoC constraints, since we are always able to find an output setpoint \((\mb{p}, \mb{q)}\) to charge or discharge the storage before reaching the apparent power limit \(\overline{\mb{s}}\).

\section{Benchmark Controller Design} \label{sec:benchmark}
To assess the performance of our OFO approach, we compare it with a benchmark controller inspired by Volt/VAR control. The benchmark controller takes the voltage deviation from the setpoint (user-defined, or nominal voltage) as one of its inputs. However, to make a fair comparison, it also takes the voltage fluctuation as the second input. The two inputs also account for deadbands and saturation:
\begin{subequations} \label{eq:droop-shape}
    \begin{align}
        \Delta \tilde{v}_{n} &= \mathrm{sgn}(v_{n, \text{prev}} - v_{n}) \nonumber \\
            & \hspace{1em} \cdot \mathcal{P}_{[0, \Delta \tilde{v}_{\tmax} - \delta/2]} \left( | v_{n, \text{prev}} - v_{n} | - \tfrac{\delta}{2} \right) \label{eqn:fluc-input} \\
        \tilde{v}_{n} &= \mathrm{sgn}(v_{\text{set}} - v_{n}) \nonumber \\
            & \hspace{1em} \cdot \mathcal{P}_{[0, \tilde{v}_{\tmax} - \delta/2]} \left( | v_{\text{set}} - v_{n} | - \tfrac{\delta}{2} \right) \label{eqn:v-input} 
    \end{align}
\end{subequations}
where \(n \in \mathcal{N} := \{ 1, \dots, N \}\) indexes all the batteries and the buses they connect to, \(\Delta\tilde{v}_{n}\) is the fluctuation signal, \(\tilde{v}_{n}\) is the voltage deviation signal, \(\delta\) is the deadband width, \(\Delta \tilde{v}_{\tmax}\) and \(\tilde{v}_{\tmax}\) are the saturated points of the two inputs, and \(v_{\text{set}}\) is a user-defined  setpoint for the voltage deviation signal.

Unfortunately, the fluctuation-related input \eqref{eqn:fluc-input} introduces essentially a derivative term into the controller (i.e., we are differencing the voltage), which could lead to instability. However, when using \eqref{eqn:v-input} alone (proportional control), the controller tends to perform poorly under frequent voltage fluctuations. One way to mitigate this issue is to redesign this benchmark controller from specifying setpoints into updating setpoints gradually over time with some small step size. This can be interpreted as integration, transforming the previous PD controller into a PI controller~\cite{franklin-1990a}. It also makes the benchmark controller more similar/comparable to the OFO controller since the OFO controller also uses an incremental update policy. Using this approach, then benchmark strategy is as follows:
\begin{subequations} \label{eq:droop-slope}
    \begin{align}
         \hat{p}_{n} &=  \hat{p}_{n,\text{prev}} + \alpha\rho \left( \frac{\Delta \tilde{v}_{n}}{\Delta \tilde{v}_{\tmax}} \theta + \frac{\tilde{v}_{n}}{\tilde{v}_{\tmax}} (1 - \theta) \right) \overline{s}_n, \\
         \hat{q}_{n} &=  \hat{q}_{n,\text{prev}} + (1 \!-\! \alpha) \rho\! \left(\! \frac{\Delta \tilde{v}_{n}}{\Delta \tilde{v}_{\tmax}} \theta + \frac{\tilde{v}_{n}}{\tilde{v}_{\tmax}} (1 \!-\! \theta) \!\!\right) \overline{s}_n,
    \end{align}
\end{subequations}
where \(\rho\) is the update step size, \(\alpha\) is the weighting factor between active power and reactive power, \(\theta\) is the weighting factor balancing the two inputs (voltage fluctuations and deviations),
and \(\overline{s}\) is the maximum apparent power limit. Combining \eqref{eq:droop-shape} and \eqref{eq:droop-slope} yields the relationships shown in Fig.~\ref{fig:droop-diagram}.
\begin{figure}
    \centering
    \includegraphics[width=1\linewidth]{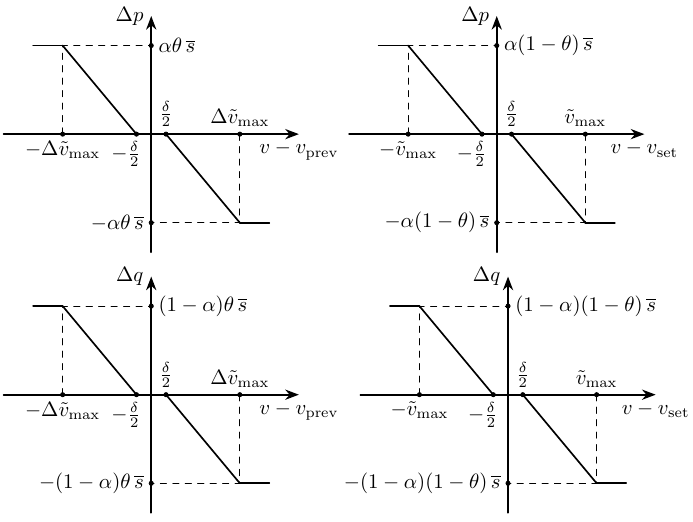}
    \caption{The piecewise-linear relationship between inputs and outputs (before projection) of the benchmark controller. The inputs (voltage fluctuations and deviations) are mapped to desired change of active and reactive power setpoints using linear control rules.}
    \label{fig:droop-diagram}
\end{figure}

After the controller updates the battery power injections based on this strategy, a sequential closed-form projection is performed to ensure the feasibility of the control decisions:
\begin{subequations}
    \begin{align}
        p_{n} &= \mathcal{P}_{[-\overline{s}_n, \overline{s}_n]} (\hat{p}_{n}), \\
        q_{n} &= \mathcal{P}_{[-\sqrt{\overline{s}_n^2 - p_{n}^2}, \sqrt{\overline{s}_n^2 - p_{n}^2}]} (\hat{q}_{n}).
    \end{align}
\end{subequations}
This controller thus serves as a comparable optimization-free benchmark for the OFO controller.

\section{Numerical Experiments} \label{sec:case-study}
In this section, we validate the proposed OFO controller using a modified IEEE 13-bus distribution system and  the real-world power consumption data from LLM inference, and compare its performance with the benchmark controller.

\subsection{Data and Settings}
The numerical experiments are carried on a modified IEEE 13-bus three-phase radial test feeder~\cite{IEEE_test}. It is relatively highly loaded with unbalanced loads, which enables us to explore the method's performance under unbalanced operation. To facilitate the experimental process and focus on the core objective of this study, a number of assumptions and simplifications are made:
\begin{enumerate}
    \item The data center loads are all balanced. Therefore, all unbalance comes from storage injections and other loads, with load data taken from the test case~\cite{IEEE_test}. Also, the data center loads operate at a fixed power factor of 0.95.
    \item The distribution network is modified by adding a few missing phases on some of the branches. Line parameters are given in Appendix~\ref{sec:system-parameter}.
    \item Additional switched capacitor banks are added to data centers buses to support local voltages. These capacitor banks are assumed to be always operating (switched on).
    \item The battery systems respond instantaneously with negligible communication, computation, and actuation delays. The control frequency is set to 10 Hz, which is also  the measurement frequency of the data center load.
\end{enumerate}
Data center load profiles are obtained from~\cite{mlenergy-neuripsdb25}. The dataset contains the GPU power measurements of different LLMs during inference. Specifically, the data corresponding to Llama-3.1 model with 4.5B parameters, 8 GPUs, and a maximum batch size of 8, since it exhibits large power fluctuations. The power measurements are scaled to the range 1--4 MW to match the load of an entire data center.

We connect three data centers at buses 611, 675, and 680, i.e., at the terminal buses, leading to significant voltage challenges. Each data center is accompanied by a battery system (with a power capacity of 500 kW per phase and an energy capacity of 200 kWh), whose parameters are given in Appendix~\ref{sec:system-parameter}. The system topology is illustrated in Fig.~\ref{fig:feeder}.
\begin{figure}
    \centering
    \includegraphics[width=1\linewidth]{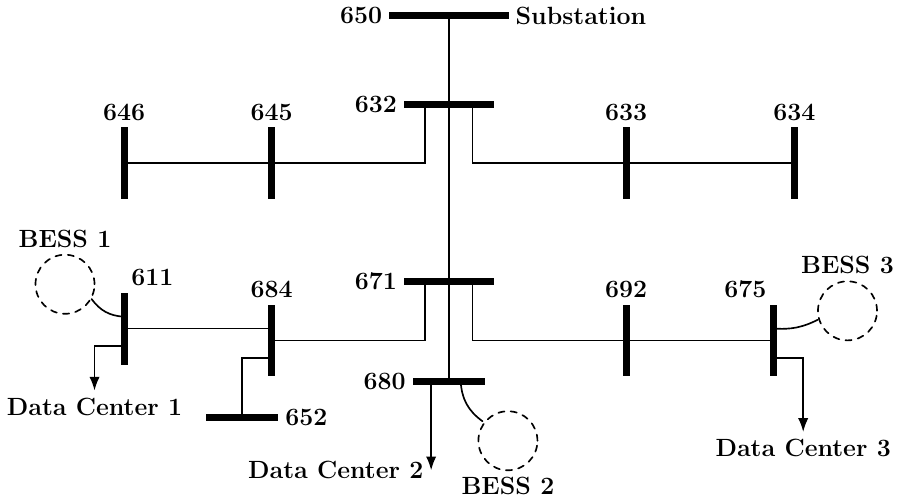}
    \caption{Topology of the IEEE 13-bus feeder used in the numerical experiments. Three data centers, each co-located with a battery system, are connected to buses 611, 675, and 680. Other loads are omitted for clarity.}
    \label{fig:feeder}
\end{figure}

\begin{figure}[t]
    \centering
    \includegraphics[width=1\linewidth]{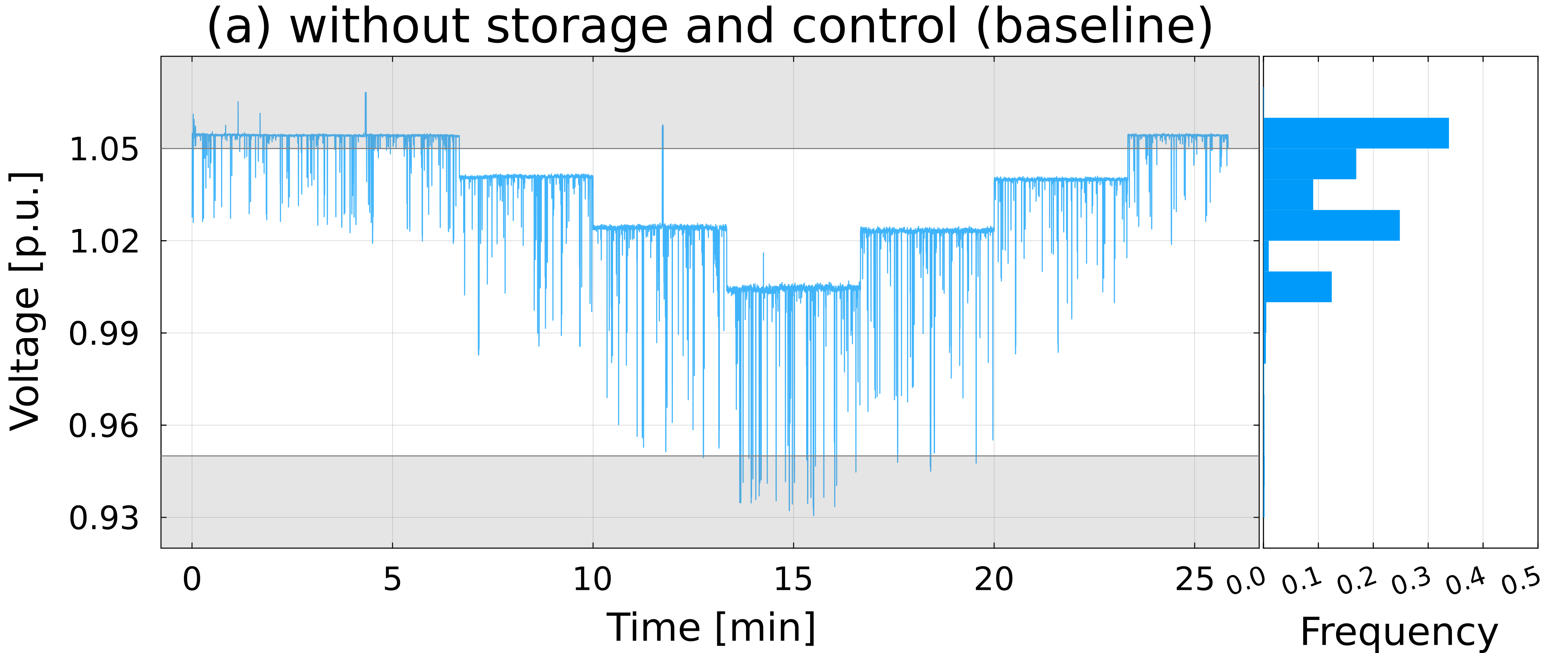} 
    
    \vspace{0.5em}
    \includegraphics[width=1\linewidth]{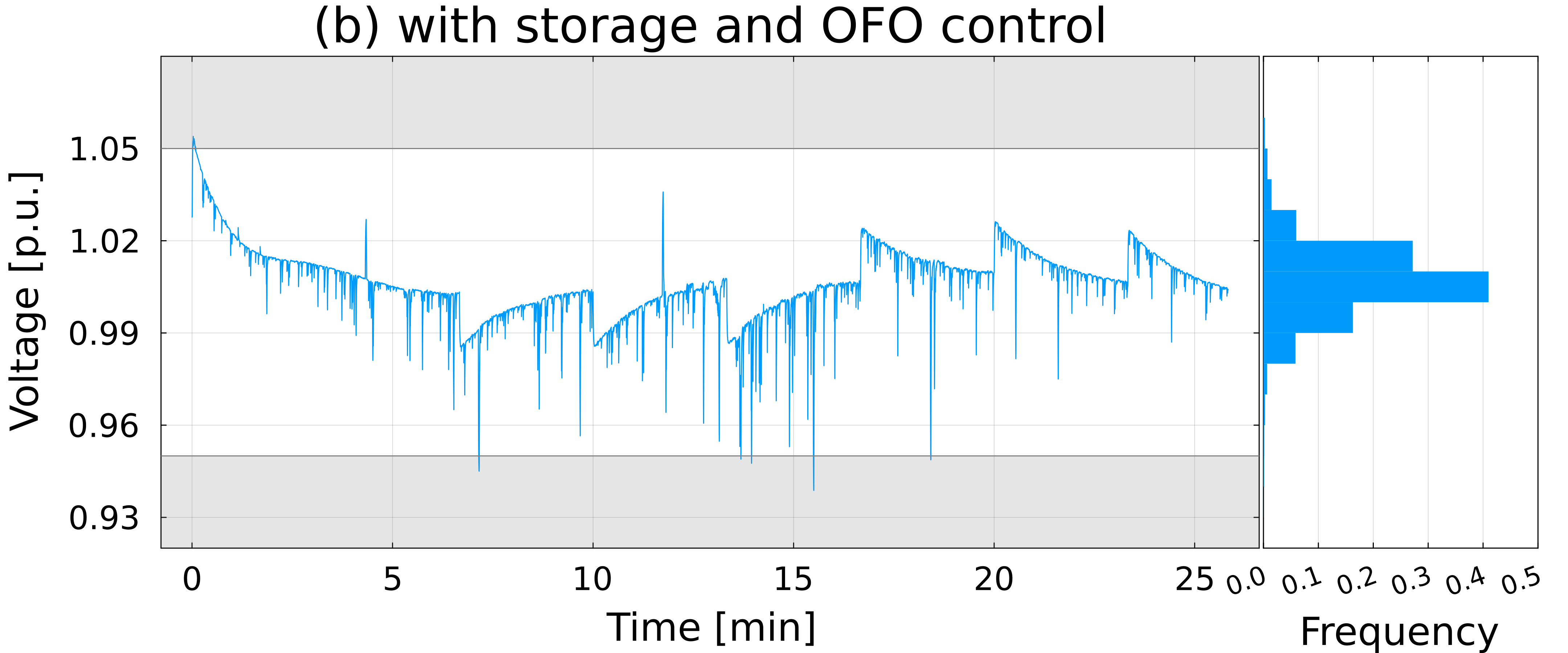}
    
    \vspace{0.5em}
    \includegraphics[width=1\linewidth]{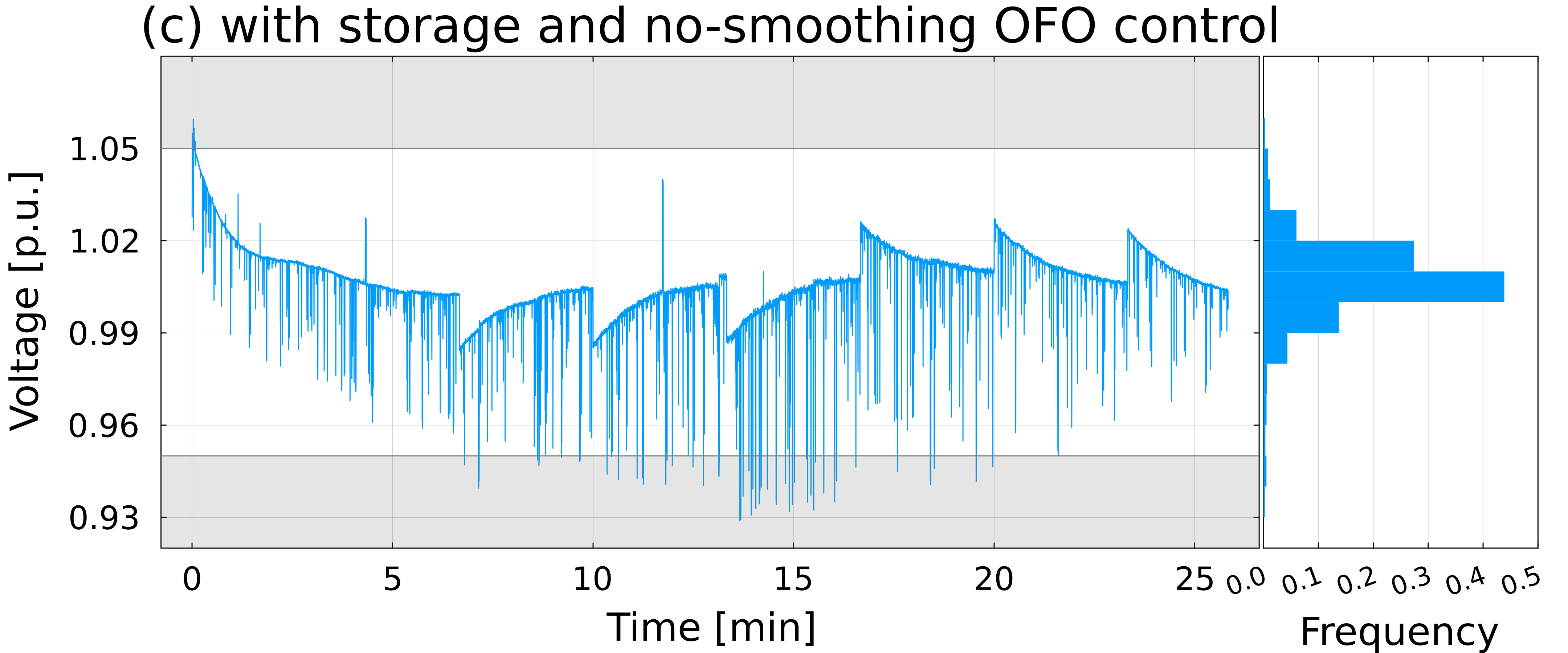}
    
    \vspace{0.5em}
    \includegraphics[width=1\linewidth]{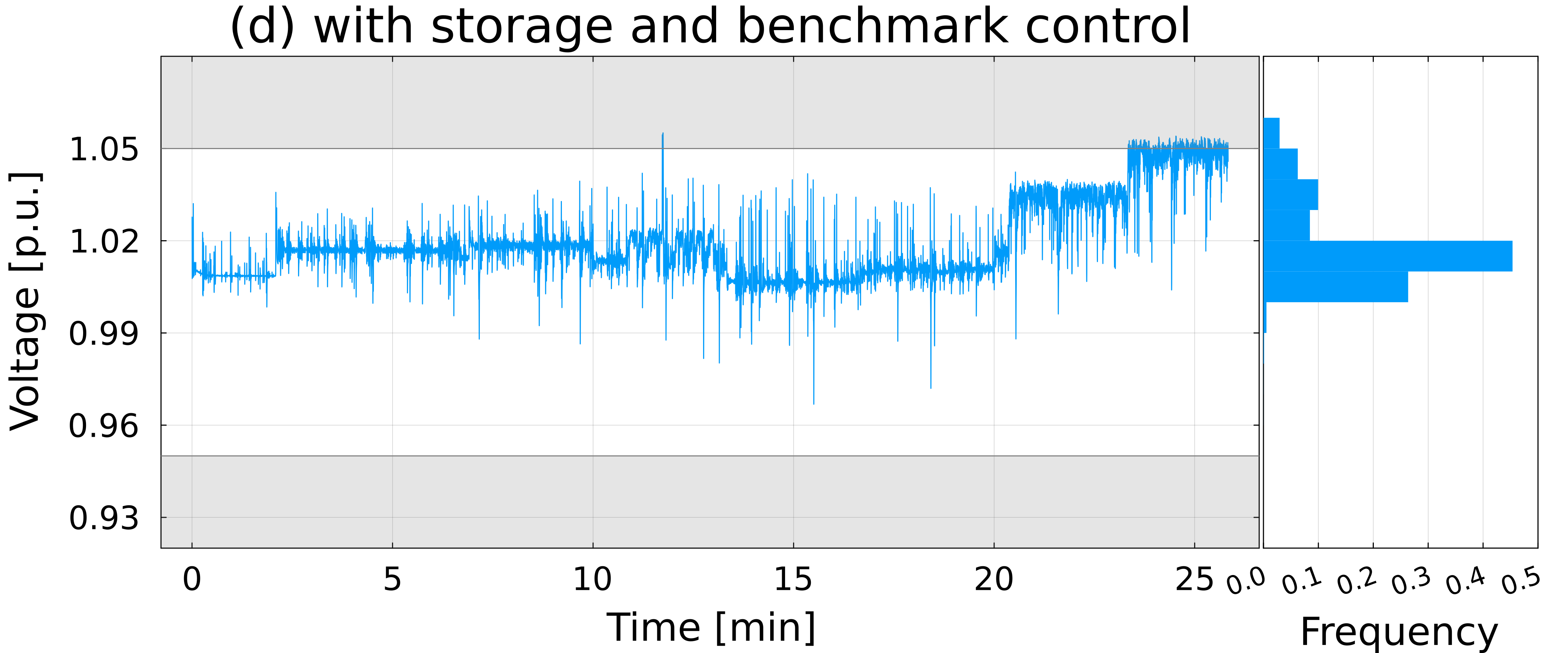}

    \caption{The voltage profiles at Bus 675 Phase c, and the corresponding histograms in different operational scenarios. (a) No control (i.e., the baseline case). (b) OFO control. (c) OFO control without the smoothing objective, i.e., \(\mb{C}_{\text{vf}} = \mb{0}\). (d) Benchmark control.}
    \label{fig:results-volt}
\end{figure}

\begin{figure}[t]
    \centering
    \includegraphics[width=1\linewidth]{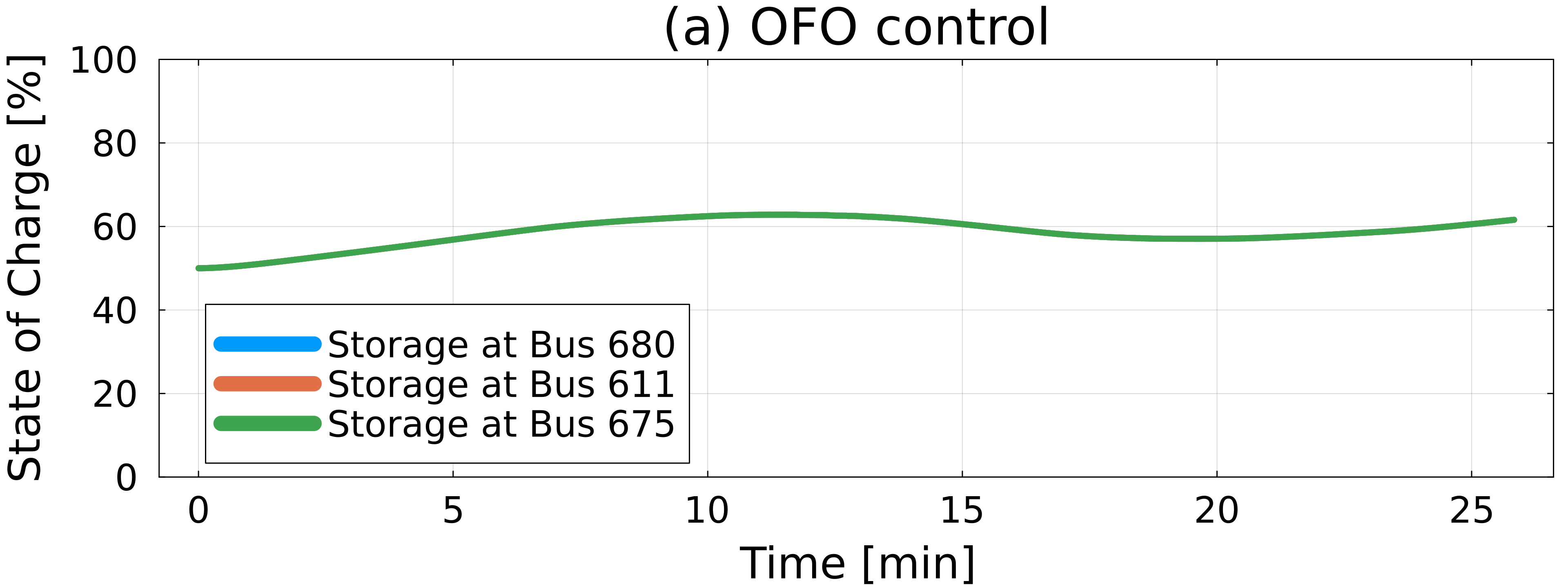}

    \vspace{0.5em}
    \includegraphics[width=1\linewidth]{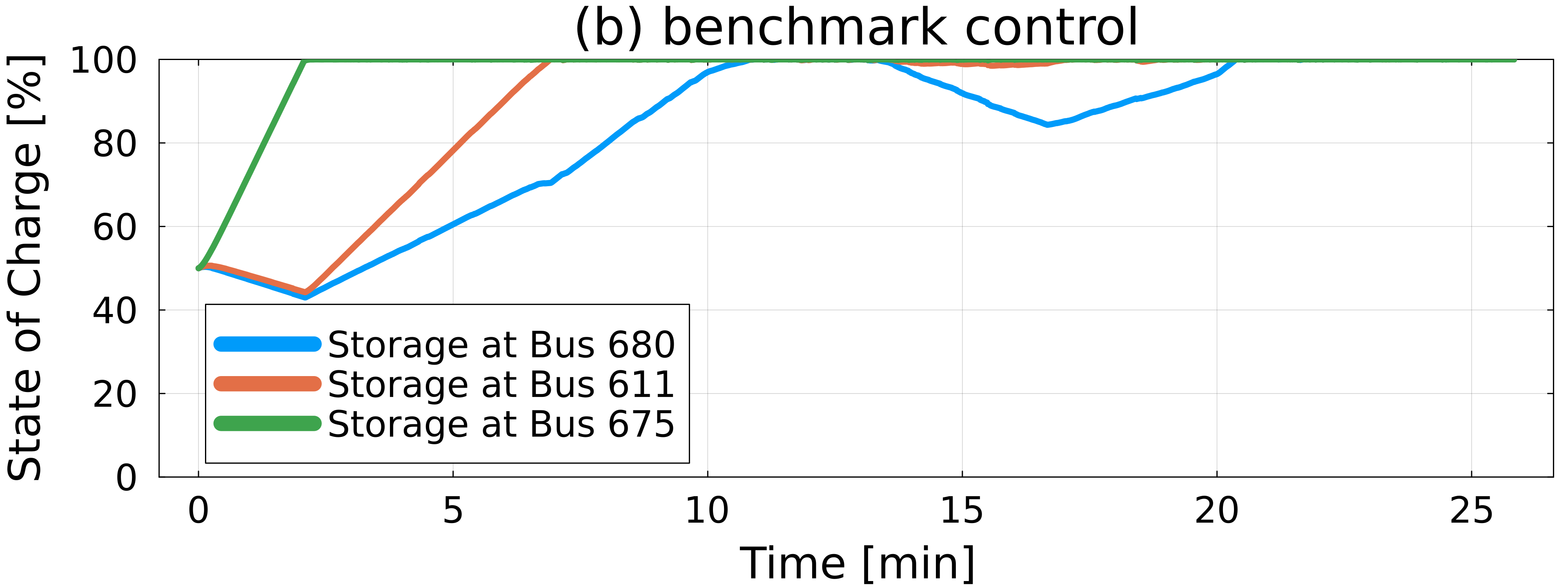}
        
    \caption{SoC dynamics for the three battery systems at buses 611, 675, and 680 under different controllers. (a) OFO control. (b) Benchmark control.}
    \label{fig:results-soc}
\end{figure}

\subsection{Results}

The voltage profile at one of the terminal buses resulting from highly-variable data center loads and no utilization of battery systems is shown in Fig.~\ref{fig:results-volt}(a). The load profile includes over- and under-voltages, which could  have adverse impacts on transformers, electric drives, and other connected equipment. The GPU power consumption also leads to a highly-variable voltage profile. The proposed OFO controller is designed to solve both problems: constraint satisfaction and voltage smoothing.

\textit{1) The OFO controller:} The  OFO controller regulates the voltage profile, as shown in Fig.~\ref{fig:results-volt}(b). It successfully smooths most fluctuations, reduces the magnitudes of voltage spikes, and manages the voltage within the safe $\pm0.05$ p.u. range. This voltage profile reduces the need for tap changing, thereby extending transformer lifetime and reducing wear and tear on expensive tap-changing equipment.

\textit{2) The OFO controller without the smoothing objective:} To illustrate the benefit of the smoothing objective, we implement the same OFO controller except with the weight on the smoothing objective \(\mb{C}_{\text{vf}}\) set to zero. The corresponding voltage profile is displayed in Fig.~\ref{fig:results-volt}(c). Although the controller is still able to bring the voltage within the voltage limits most of the time, voltage fluctuations lead to persistent violations of the lower voltage limit, thus negatively affecting transformers and machines connected to the feeder. In contrast, the OFO with the smoothing objective does not have this problem. 

\textit{3) The benchmark controller:} The results of the benchmark controller are reported in Fig.~\ref{fig:results-volt}(d). The benchmark controller (with parameters $\alpha=\theta=0.5$) brings the voltage within limits and suppresses large spikes at the beginning. However, this effect does not last long, and unlike under the OFO controller, the upper-voltage limit is violated starting from $t=23$ min. Fig.~\ref{fig:results-soc} shows the utilization of the three battery systems, which explains the failure of the benchmark controller to deliver consistent performance for the entire duration of the simulation. Observe that the OFO controller maintains the SoC level at roughly 50\% for all three battery systems throughout the simulation period, while the benchmark controller gradually hits the upper SoC limit of all three systems, leading to the step changes in the voltage profile in Fig.~\ref{fig:results-volt}(d) until the voltage exceeds the upper limit. This difference is inherent to the controller design: the OFO controller optimizes the cost of control effort and accounts for the SoC, while the benchmark controller is ignorant of both the control cost and SoC constraints. As a result, the benchmark controller aggressively pursues the control goals initially but quickly depletes the storage reserve.

Table~\ref{tab:numerical-metrics} compares all controllers in terms of voltage variance, voltage range, and accumulated voltage limit violation magnitude. The OFO controller achieves the minimum voltage variance and accumulated violation magnitude. Removing the smoothing objective from OFO increases the voltage range, and large dips contribute to the accumulated violation exceeding that of the benchmark controller, hence justifying the smoothing objective in the OFO controller design.

\begingroup
\begin{table}
    \caption{Numerical Performance Evaluations of Different Controller}
    \label{tab:numerical-metrics}
\begin{tabular}{ccccc}
    \toprule
    \multirow{2}{*}{{Voltage}} & \multicolumn{4}{c}{{Controller}}  \\
    \cmidrule(lr){2-5}
    {Metrics}\(^*\) & \multirow{2}{*}{{Baseline}} & \multirow{2}{*}{{OFO}} & {OFO w/o} & \multirow{2}{*}{{Benchmark}} \\
     & & & {smoothing} & \\
    \midrule
    Variance & \(3.69 {\times10}^{-4}\) & \(9.62 {\times10}^{-5}\) & \(1.35 {\times10}^{-4}\) & \(1.74 {\times10}^{-4}\) \\
    \addlinespace
    Range & 0.138 & 0.097 & 0.111 & 0.088 \\
    \addlinespace
    Total voltage & \multirow{2}{*}{15.60} & \multirow{2}{*}{0.076} & \multirow{2}{*}{1.029} & \multirow{2}{*}{0.676} \\
    violation\(^{**}\) & & & &\\
    \bottomrule \addlinespace[0.35em]
    \multicolumn{5}{l}{\(^*\)All in p.u. and computed using data points after first 2 min to exclude} \\
    \multicolumn{5}{l}{~ initial dynamics.} \\
    \multicolumn{5}{l}{\(^{**}\)Voltage magnitude violation accumulated during the control period.}
\end{tabular}
\end{table}
\endgroup

\section{Conclusion} \label{sec:conclusion}
We have developed an OFO voltage controller that leverages distributed battery systems to address the strain posed by large-scale data centers connected to distribution feeders. Our experiments on an IEEE test feeder show that the OFO controller is capable of smoothing the voltage impact of the real GPU power traces from inference on Llama-3.1 model. Unlike local controllers ignorant of storage constraints and quickly depleting the storage resource, the OFO controller maintains the SoC on regular levels while resolving voltage constraint violations and smoothing the voltage profile. The smoothing objective is instrumental: Removing the smoothing objective leads to persistent lower-voltage violations despite nominal constraint enforcement, increasing accumulated violations and hence the risk to equipment connected to the distribution feeder. The statistical analysis revealed that OFO minimizes voltage variance and accumulated violation magnitude; without smoothing, its performance degrades to benchmark levels, validating the controller design choices.

\section{AI Usage Disclosure}

The authors used AI for polishing the narrative (spell-checking and streamlining). None of the narrative was directly produced by AI; it was solely used to advise the authors, not to replace them. The model and ideas are the intellectual contribution of the authors, and no AI model participated in developing this intellectual contribution.

\appendix 

\subsection{Grid Modeling} \label{sec:grid-model}
The compact form of LinDist3Flow \cite{taheri_optimizing_2025}, the 3-phase extension of the LinDistFlow first proposed in~\cite{baran_network_1989}, is:
\begin{align}
    \mb{v} = \mb{v}_\nom + \underbrace{\mb{F} \mathrm{bdiag}(\mb{H}^\text{P}) \mb{F}^\top}_{\mb{R}_{3\phi}} \mb{p} + \underbrace{\mb{F} \mathrm{bdiag}(\mb{H}^\text{Q}) \mb{F}^\top}_{\mb{X}_{3\phi}} \mb{q},
\end{align}
where \(\mb{v} = [v_{1,a}^2 \quad v_{1,b}^2 \quad v_{1,c}^2 \quad \cdots \quad v_{n,a}^2 \quad v_{n,b}^2 \quad v_{n,c}^2]^\top\) is the vector of squared voltages across all buses and all phases; \(\mb{v}_\nom\) is the vector of nominal voltages across all buses; \(\mb{F}\) is the inverse of the 3-phase network incidence matrix; \(\mathrm{bdiag}(\cdot)\) is the block-diagonal operator that creates a block diagonal matrix based on input; \(\mb{p} = [p_{1,a} \quad p_{1,b} \quad p_{1,c} \quad \cdots \quad p_{n,a} \quad p_{n,b} \quad p_{n,c}]^\top, \mb{q} = [q_{1,a} \quad q_{1,b} \quad q_{1,c} \quad \cdots \quad q_{n,a} \quad q_{n,b} \quad q_{n,c}]^\top\) are vectors of phase-wise active and reactive power injections; and \(\mb{H}\) are 3-phase impedance matrices:
{\setlength{\arraycolsep}{0.325em}
\begin{subequations}
\begin{align}
    \hspace{-0.5em}\mb{H}^\text{P} &=\! \begin{bmatrix}
        2r_{aa} & -r_{ab} \!+\! \sqrt{3} x_{ab} & -r_{ac} \!-\! \sqrt{3} x_{ac} \\
        -r_{ba} \!-\! \sqrt{3} x_{ba} & 2r_{bb} & -r_{ba} \!+\! \sqrt{3} x_{ba} \\
        -r_{ca} \!+\! \sqrt{3} x_{ca} & -r_{cb} \!-\! \sqrt{3} x_{cb} & 2r_{cc} 
    \end{bmatrix}\!,\! \\
    \hspace{-0.5em}\mb{H}^\text{Q} &=\! \begin{bmatrix}
        2x_{aa} & -x_{ab} \!-\! \sqrt{3} r_{ab} & -x_{ac} \!+\! \sqrt{3} r_{ac} \\
        -x_{ba} \!+\! \sqrt{3} r_{ba} & 2x_{bb} & -x_{ba} \!-\! \sqrt{3} r_{ba} \\
        -x_{ca} \!-\! \sqrt{3} r_{ca} & -x_{cb} \!+\! \sqrt{3} r_{cb} & 2x_{cc} 
    \end{bmatrix}\!.\!
\end{align}
\end{subequations}
}

Then, a linearized approximation of \eqref{eq:master-grid-model} can be expressed as:
\begin{equation}
    \mb{v}_t = \mb{v}_{\nom} + \mb{R}_{3\phi} (\mb{p}_t \!-\! \mb{d}_t^{\text{P}}) + \mb{X}_{3\phi} (\mb{q}_t - \mb{d}_t^{\text{Q}}), \quad \forall t \in \mathcal{T}, \label{eq:LDF}
\end{equation}
where \( \mb{d}_t = \mb{d}_t^{\text{P}} + j \mb{d}_t^{\text{Q}}\).

\subsection{Storage Modeling} \label{sec:storage-model}
The SoC variation is of key importance for storage system models, especially when accounting for storage physical constraints. To better capture the dynamic of SoC, our model incorporates power electronics converter efficiency. Notably, charging and discharging processes have different efficiency. The SoC varies based on how much energy the system absorbs and outputs -- system SoC decreases when generating active power, and increases when absorbing active power. Importantly, due to inherent converter losses, the system also discharges energy even when providing reactive power alone.

To quantify the active power loss associated with the draw of reactive power, we express the relationship between instantaneous power \(p(t)\), voltage \(v(t)\) and current \(i(t)\):
\begin{equation}
    \begin{split}
        p(t) = v(t) i(t) = &\underbrace{V_\rms I_\rms \cos (\phi)}_P \cdot \big( 1 + \cos (2 \omega t) \big) \\
        &- \underbrace{V_\rms I_\rms \sin (\phi)}_Q \cdot \sin (2 \omega t), \label{eq:p-vi}    
    \end{split}
\end{equation}
where \(\phi\) is the power factor and \(\omega\) is the grid frequency. The term containing \(\sin(2 \omega t)\) represents reactive power, meaning the system is continuously exchanging energy back and forth with the external circuit. Due to converter inefficiency, the system needs to supply more energy and receive less energy to ensure the external demand is satisfied, resulting in energy loss during reactive power transmission. 
The absolute difference between the actual instantaneous power and its setpoint comprises the loss related to reactive power, as demonstrated in Fig.~\ref{fig:Q-loss}. The grid-side power is set to be purely sinusoidal, based on which the storage-side power is computed. The instantaneous power loss is expressed piece-wise:
\begin{align}
    &p_{\text{loss,re}}(t)= \nonumber \\
    &\hspace{1.5em} \begin{cases}
        \Big| Q_{\text{set}} \sin (2 \omega t) \cdot (\frac{1}{\eta^+} - 1) \Big|, t \in [0, \frac{T}{4}) \cup [\frac{T}{2}, \frac{3T}{4}) \\
        \Big| Q_{\text{set}} \sin(2 \omega t) \cdot (1 - \eta^-) \Big|, t \in [\frac{T}{4}, \frac{T}{2}) \cup [\frac{3T}{4}, T)
    \end{cases} \hspace{-1em} ,
    \label{eq:p_loss,re}
    \vspace{1em}
\end{align}
where \(Q_\text{set} = V_{\rms,\text{set}} I_{\rms,\text{set}} \sin (\phi_\text{set})\) is the reactive power setpoint measurable by grid, \(\eta^+\) and \(\eta^-\) are discharging and charging efficiency.

\begin{figure}
    \centering
    \includegraphics[width=\linewidth]{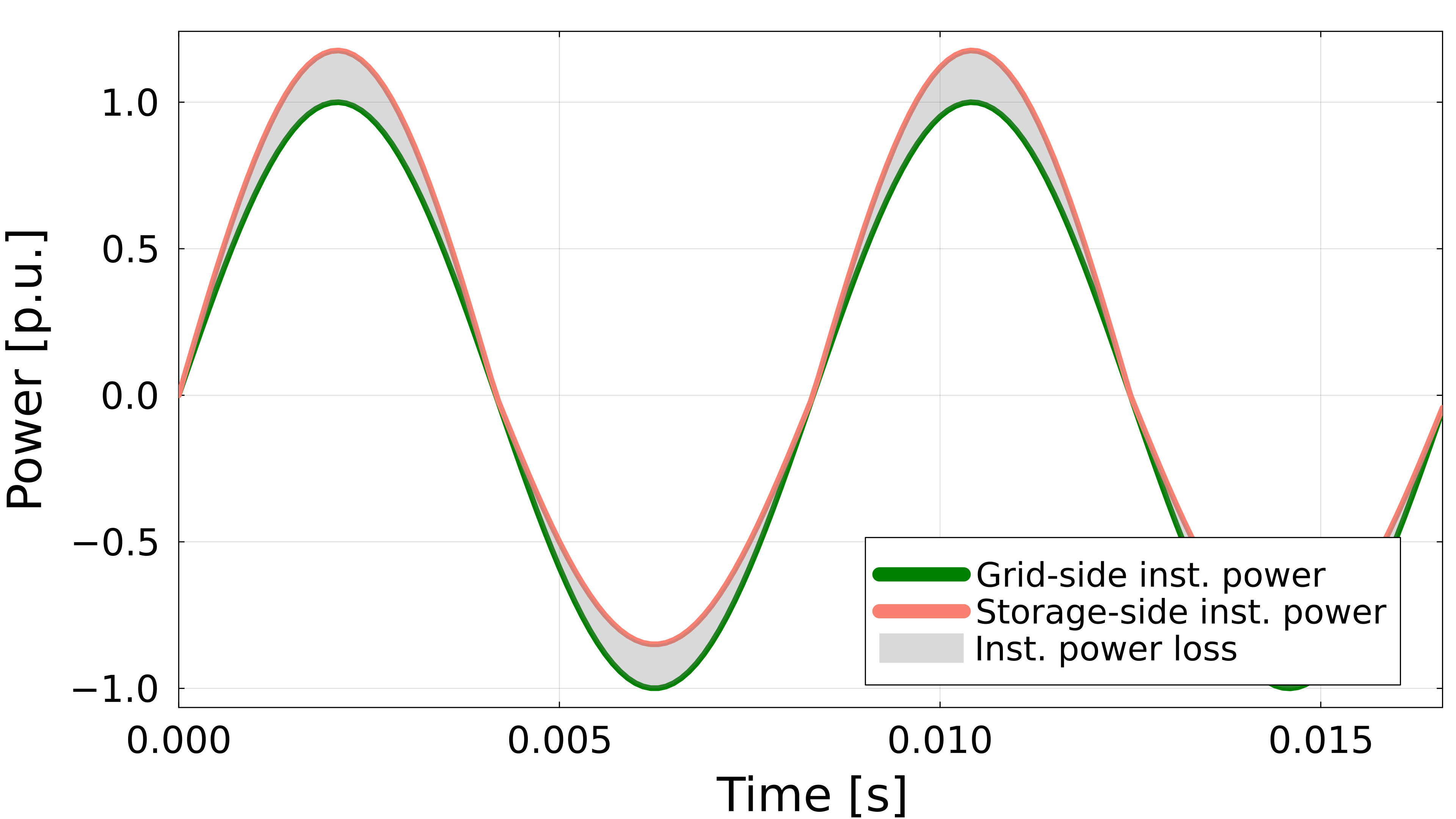}
    \caption{The active power loss related to reactive power. The green curve is the instantaneous power that grid exchanges with the storage system when transmitting reactive power (i.e., grid reference setpoint), positive indicating charging. The orange curve is the actual instantaneous power throughput of energy storage unit. Due to loss, storage absorb less power while discharging and inject more power while charging. Between the two curves is the instantaneous power loss of the converter.}
    \label{fig:Q-loss}
\end{figure}

Given that the distribution network is running at a relatively high frequency (i.e., 60 Hz), we presume the reactive power setpoint remains unchanged over a single cycle. Also, the truncation of one cycle do not contribute much to the average power in a long time period. Thus, the average power of loss can be calculated, as is shown in \eqref{eq:Q-loss}.
\begin{align}
    \bar{P}_{\text{loss,re}} &= \frac{1}{T} \int_0^T p_{\text{loss,re}}(t) \dif t \nonumber\\
    &= \frac{2}{T} \bigg( \int_0^\frac{T}{4} \bigg| q \sin (2 \omega t) \cdot (\frac{1}{\eta^+} - 1) \bigg| \dif t \nonumber\\
    &\hspace{3em} + \int_\frac{T}{4}^\frac{T}{2} \bigg| q \sin (2 \omega t) \cdot (1 - \eta^-) \bigg| \dif t \bigg) \nonumber\\
    &= \frac{1 - \eta^+ \eta^-}{\pi \eta^+ } | q |. \label{eq:Q-loss}
\end{align}

Thus, a model of SoC transition equation \eqref{eq:master-socdynm} involving both active power and reactive power within time step \(t\) can be derived:
\begin{align}
    \mb{e}_{t+1} = \mb{e}_{t} + \Big(\frac{1}{\eta^-} \mb{p}_{t}^- \Delta t & - \eta^+ \, \mb{p}_{t}^+ \Delta t \nonumber\\
    &- \frac{1 - \eta^+ \eta^-}{\pi \eta^+} (\mb{q}_t^+ + \mb{q}_t^-) \Delta t \Big). \label{eq:SoC}
\end{align}
where the split of \(\mb{p}^+, \mb{q}^+\) and \(\mb{p}^-, \mb{q}^-\) follows the relationship:%
\begin{subequations}
    \begin{align}
        & \mb{p} = \mb{p}^+ - \mb{p}^-, \quad 0 < \mb{p}^+ \perp \mb{p}^- > 0, \label{eq:off-p+p-} \\
        & \mb{q} = \mb{q}^+ - \mb{q}^-, \quad 0 < \mb{q}^+ \perp \mb{q}^- > 0 .\label{eq:off-q+q-}
    \end{align}
\end{subequations}

The loss caused by reactive power is present as the last term in \eqref{eq:SoC}. Although it will be subtly different when active and reactive power are present at the same time, the difference will be very small given \(\eta^+\) and \(\eta^-\) are close. This reactive loss term adds new dynamics into system SoC variation, preventing storage systems from persistent reactive power exchanging at low SoC.

\balance
\subsection{Line Parameters} \label{sec:system-parameter}
All lines in the test feeder are assumed to be overhead lines of configuration 602, with the following resistance and reactance matrices:
    \begin{subequations}
    \begin{align}
        &\mb{R}_{602} = \begin{bmatrix}
            0.7526 & 0.1580 & 0.1560 \\
             & 0.7475 & 0.1535 \\
             & & 0.7436
        \end{bmatrix} \Omega / \text{mi},& \\
        &\mb{X}_{602} = \begin{bmatrix}
            1.1814 & 0.4236 & 0.5017 \\
             & 1.1983 & 0.3849 \\
             & & 1.2112
        \end{bmatrix} \Omega / \text{mi}.&
    \end{align}
    \end{subequations}
    
\bibliographystyle{IEEEtran}
\bibliography{references}

@article{li_unseen_2024,
	title = {The {Unseen} {AI} {Disruptions} for {Power} {Grids}: {LLM}-{Induced} {Transients}},
	doi = {10.48550/arXiv.2409.11416},
	urldate = {2025-01-30},
	publisher = {arXiv},
	author = {Li, Yuzhuo and Mughees, Mariam and Chen, Yize and Li, Yunwei Ryan},
	month = sep,
    journal = {arXiv preprint arXiv:2409.11416},
	year = {2024},
}

@article{baran_network_1989,
	title = {Network reconfiguration in distribution systems for loss reduction and load balancing},
	volume = {4},
	issn = {1937-4208},
	doi = {10.1109/61.25627},
	number = {2},
	journal = {IEEE Transactions on Power Delivery},
	author = {Baran, M.E. and Wu, F.F.},
	month = apr,
	year = {1989},
	pages = {1401--1407},
}

@article{taheri_optimizing_2025,
	title = {Optimizing {Parameters} of the {LinDistFlow} {Power} {Flow} {Approximation} for {Distribution} {Systems}},
	doi = {10.48550/arXiv.2404.05125},
    journal = {arXiv preprint arXiv:2404.05125},
	publisher = {arXiv},
	author = {Taheri, Babak and Gupta, Rahul K. and Molzahn, Daniel K.},
	month = may,
	year = {2025},
}

@misc{hauswirth_optimization_2024,
	title = {Optimization {Algorithms} as {Robust} {Feedback} {Controllers}},
	doi = {10.48550/arXiv.2103.11329},
	publisher = {arXiv},
	author = {Hauswirth, Adrian and He, Zhiyu and Bolognani, Saverio and Hug, Gabriela and Dörfler, Florian},
	month = jan,
	year = {2024},
    journal = {arXiv preprint arXiv:2103.11329},
}

@article{ortmann_experimental_2020,
	title = {Experimental validation of feedback optimization in power distribution grids},
	volume = {189},
	issn = {03787796},
	doi = {10.1016/j.epsr.2020.106782},
	language = {en},
	journal = {Electric Power Systems Research},
	author = {Ortmann, Lukas and Hauswirth, Adrian and Caduff, Ivo and Dörfler, Florian and Bolognani, Saverio},
	month = dec,
	year = {2020},
	pages = {106782},
}

@electronic{web_connectedsolutions,
	key = {{ConnectedSolutions}},
    title = {{ConnectedSolutions} {\textbar} {National} {Grid}},
	url = {https://www.nationalgridus.com/MA-Home/Connected-Solutions/BatteryProgram},
	urldate = {2025-11-25},
}

@electronic{web_wattsmart,
    key = {{Wattsmart}},
	title = {Wattsmart {Battery} program},
	url = {https://www.rockymountainpower.net/savings-energy-choices/wattsmart-battery-program.html},
	urldate = {2025-11-25},
}

@electronic{web_BYOD,
    key = {{Green Mountain Power}},
	title = {Bring {Your} {Own} {Device} - {Green} {Mountain} {Power}},
	url = {https://greenmountainpower.com/rebates-programs/home-energy-storage/bring-your-own-device/},
	urldate = {2025-11-25},
}

@inproceedings{mlenergy-neuripsdb25,
    title={The {ML.ENERGY} Benchmark: Toward Automated Inference Energy Measurement and Optimization},
    author={Jae-Won Chung and Jeff J. Ma and Ruofan Wu and Jiachen Liu and Oh Jun Kweon and Yuxuan Xia and Zhiyu Wu and Mosharaf Chowdhury},
    year={2025},
    booktitle={NeurIPS Datasets and Benchmarks},
}

@electronic{IEA_energy_2025,
    author = {IEA},
	title = {Energy and {AI}},
	url = {https://www.iea.org/reports/energy-and-ai},
	urldate = {2025-11-27},
	journal = {IEA},
	month = apr,
	year = {2025}
}

@electronic{nerc_incident_2025,
    author = {NERC},
	title = {Incident Review {\textbar} Considering Simultaneous Voltage-Sensitive Load Reductions},
	url = {https://www.nerc.com/globalassets/our-work/reports/event-reports/incident_review_large_load_loss.pdf},
	urldate = {2025-10-08},
	journal = {NERC},
	month = jan,
	year = {2025}
}

@electronic{IEEE_test,
  author = {{IEEE Distribution System Analysis Subcommittee}},
  title = {{IEEE PES Test Feeder}},
  url = {https://cmte.ieee.org/pes-testfeeders/},
  year = {1991}
}

@article{mitchell-jackson_data_2003,
	title = {Data center power requirements: measurements from {Silicon} {Valley}},
	volume = {28},
	copyright = {https://www.elsevier.com/tdm/userlicense/1.0/},
	issn = {03605442},
	shorttitle = {Data center power requirements},
	doi = {10.1016/S0360-5442(03)00009-4},
	language = {en},
	number = {8},
	journal = {Energy},
	author = {Mitchell-Jackson, J. and Koomey, J.G. and Nordman, B. and Blazek, M.},
	month = jun,
	year = {2003},
	pages = {837--850},
}

@misc{choukse_power_2025,
	title = {Power {Stabilization} for {AI} {Training} {Datacenters}},
	doi = {10.48550/arXiv.2508.14318},
	publisher = {arXiv},
	month = aug,
	year = {2025},
	journal = {arXiv preprint arXiv:2508.14318},
	author = {Choukse, Esha and others},
}

@inproceedings{feehally_battery_2016,
	title = {Battery energy storage systems for the electricity grid: {UK} research facilities},
	doi = {10.1049/cp.2016.0257},
	booktitle = {8th {IET} {International} {Conference} on {Power} {Electronics}, {Machines} and {Drives} ({PEMD} 2016)},
	author = {Feehally, T and Forsyth, A J and Todd, R and Foster, M P and Gladwin, D and A Stone, D and Strickland, D},
	month = apr,
	year = {2016},
	pages = {1--6},
}

@book{franklin-1990a,
    author = {Franklin, G. F. and Powell, J. D. and Workman, M. L.},
    edition = {Third},
    publisher = {Ellis-Kagle Press},
    title = {Digital Control of Dynamic Systems},
    year = 1998,
    pages = {66},
}

@misc{dugan_opendss,
    author = {Dugan, R. C. and Montenegro, D.},
	title = {{The Open Distribution System Simulator (OpenDSS) Reference Guide}},
    address = {Electric Power Research Institute. Washington, DC: Electric Power Research Institute},
	year = 2020,
}

@misc{msenergy_tapchanger,
    key = {msenergy},
    title = {{Transformer Electrical Components Maintenance – LoadTap Changers}},
    url = {https://msenergy.com/transformer-electrical-maintenance/}
}

@article{ai-grid-impact-arxiv25,
  title={Electricity Demand and Grid Impacts of AI Data Centers: Challenges and Prospects}, 
  author={Xin Chen and Xiaoyang Wang and Ana Colacelli and Matt Lee and Le Xie},
  year={2025},
  journal={arXiv preprint arXiv:2509.07218},
}

@article{mieth2018data,
  title={Data-driven distributionally robust optimal power flow for distribution systems},
  author={Mieth, Robert and Dvorkin, Yury},
  journal={IEEE Control Systems Letters},
  volume={2},
  number={3},
  pages={363--368},
  year={2018},
  publisher={IEEE}
}

@ARTICLE{8463580,
  author={Guo, Yifei and Wu, Qiuwei and Gao, Houlei and Chen, Xinyu and Østergaard, Jacob and Xin, Huanhai},
  journal={IEEE Transactions on Sustainable Energy}, 
  title={MPC-Based Coordinated Voltage Regulation for Distribution Networks With Distributed Generation and Energy Storage System}, 
  year={2019},
  volume={10},
  number={4},
  pages={1731-1739},
  doi={10.1109/TSTE.2018.2869932}}

@article{bernstein2019real,
  title={Real-time feedback-based optimization of distribution grids: A unified approach},
  author={Bernstein, Andrey and Dall’Anese, Emiliano},
  journal={IEEE Transactions on Control of Network Systems},
  volume={6},
  number={3},
  pages={1197--1209},
  year={2019},
  publisher={IEEE}
}

@article{colot2024optimal,
  title={Optimal power flow pursuit via feedback-based safe gradient flow},
  author={Colot, Antonin and Chen, Yiting and Corn{\'e}lusse, Bertrand and Cort{\'e}s, Jorge and Dall’Anese, Emiliano},
  journal={IEEE Transactions on Control Systems Technology},
  year={2024},
  publisher={IEEE}
}

@INPROCEEDINGS{astrom_windup,
  author={Astrom, Karl Johan and Rundqwist, Lars},
  booktitle={1989 American Control Conference}, 
  title={Integrator Windup and How to Avoid It}, 
  year={1989},
  volume={},
  number={},
  pages={1693-1698},
  doi={10.23919/ACC.1989.4790464}}

@techreport{garg2025considerations,
    author = {Garg, Aditie and others},
    title = {Considerations for Distributed Edge Data Centers and Use of Building Loads to Support Large Interconnections},
    institution = {NREL},
    year = {2025}
}
\balance

\endgroup
\end{document}